\documentclass{jpsj-suppl}
\usepackage{color}
\renewcommand{\vec}[1]{\mbox{\boldmath $#1$}}

\title{
Two-nucleon Correlations in the Decay of Unbound Nuclei beyond the 
Drip Lines}

\author{K. \textsc{Hagino}$^{1,2}$ and H. \textsc{Sagawa}$^{3,4}$}

\inst{$^1$
Department of Physics, Tohoku University, Sendai 980-8578,  Japan \\
$^2$
Research Center for Electron Photon Science, Tohoku University, 1-2-1 Mikamine, Sendai\\ ~~982-0826, Japan \\
$^3$
Center for Mathematics and Physics,  University of Aizu,
Aizu-Wakamatsu, Fukushima 965-8560,\\ ~~Japan \\
$^4$
RIKEN Nishina Center, Wako 351-0198, Japan}

\abst{We discuss the two-neutron emission decay of $^{26}$O nucleus 
from a viewpoint of dineutron correlations between the valence neutrons. 
We first discuss how the dineutron correlation is realized both in the 
coordinate space and in the momentum space by employing a three-body model.  
We then discuss the decay energy spectrum as well as 
the angular distribution of emitted neutrons. 
We show that 
an emission of two neutrons 
in back-to-back directions is enhanced, which we interpret as 
a clear manifestation of the dineutron correlation between the valence 
neutrons. 
}

\kword{dineutron correlations, unbound nuclei, three-body model}

\begin{document}
\maketitle

\section{Introduction}

In recent years, the dineutron and diproton correlations 
have attracted lots of attention in connection with physics 
of weakly bound nuclei\cite{Matsuo05,Matsuo06,HS05,PSS07}. 
These are spatial correlations with which two nucleons are 
localized in the surface region of nuclei. 
The mean opening angles between the valence neutrons 
with respect to the center of the core
of Borromean nuclei 
have been extracted from the measured 
Coulomb breakup cross sections\cite{N06,A99} and found to be 
significantly smaller than the value 
for the independent neutrons,  
90 degrees\cite{N06,HS07,BH07}. 
The extracted values are 
$\langle \theta_{12}\rangle = 65.2\pm 12.2$ degrees for
$^{11}$Li and $74.5\pm 12.1$ degrees for $^6$He \cite{HS07}, 
clearly indicating the existence of the dineutron correlation in 
these nuclei. 

As other probes for the dineutron and diproton correlations, 
a two-proton radioactivity, that is, 
a spontaneous emission of two protons from 
proton-unbound nuclei, has been considered to be a good candidate 
\cite{PKGR12}. 
Very recently, the ground state {\it two-neutron} emissions 
have also been observed, {\it e.g.,} in $^{16}$Be \cite{Spyrou12}, 
$^{13}$Li \cite{Kohley13}, and $^{26}$O \cite{Lunderberg12,Caesar13}. 
An attractive feature of these phenomena is that the two valence nucleons 
are emitted directly from the ground state in contrast to the 
Coulomb breakup, in which a nucleus is first excited by the external electromagnetic 
interaction. 

In this contribution, we present three-body model calculations 
for the two-neutron 
decay of the $^{26}$O nucleus \cite{HS14,HS14b}. 
To this end, we assume
the three-body structure of $^{24}$O+$n$+$n$ and 
take into account the couplings to 
continuum by the Green's function technique, 
which was originally invented in order to describe 
the continuum dipole excitations of $^{11}$Li \cite{EB92}. 

We shall discuss a role of dineutron correlation in the decay process, and thus 
let us first discuss 
the mechanism of the dineutron correlation in the next section. 

\section{Dineutron correlation in the coordinate and in the momentum spaces}

As has been discussed in Refs. \cite{PSS07,CIMV84,HVPS11}, 
the dineutron correlation, that is, the spatial localization 
of the two-particle density, is caused by a coherent admixture of 
many configurations with opposite parity states. 
Symbolically, let us write a two-particle wave function as,
\begin{equation}
\Psi(\vec{r},\vec{r}')=\alpha\, \Psi_{\rm ee}(\vec{r},\vec{r}')
+\beta\, \Psi_{\rm oo}(\vec{r},\vec{r}'),
\label{twowf}
\end{equation}
where $\Psi_{\rm ee}$ and $\Psi_{\rm oo}$ are two-particle wave functions 
with even and odd angular momentum states, respectively. 
Notice that $\Psi_{\rm ee}(\vec{r},\vec{r}')=\Psi_{\rm ee}(\vec{r},-\vec{r}')$ 
and $\Psi_{\rm oo}(\vec{r},\vec{r}')=-\Psi_{\rm oo}(\vec{r},-\vec{r}')$. 
For nuclear systems, the coefficients $\alpha$ and $\beta$ are such that 
the interference term in the two-particle density, 
$\alpha^*\beta\Psi^*_{\rm ee}\Psi_{\rm oo}+c.c.$, is positive 
for $\vec{r}'=\vec{r}$ while it is negative for 
$\vec{r}'=-\vec{r}$ \cite{HVPS11}. 
That is, the two-particle density is enhanced for the nearside configuration 
with $\vec{r}\sim\vec{r}'$ as compared to the farside configuration 
with $\vec{r}\sim-\vec{r}'$,  provided that the wave functions
$\Psi_{\rm ee}(\vec{r},\vec{r}')$ and $\Psi_{\rm oo}(\vec{r},\vec{r}')$ are mixed coherently. 
This is nothing but the dineutron correlation. 

Now let us take the Fourier transform of $\Psi(\vec{r},\vec{r}')$, 
that is, 
\begin{equation}
\widetilde{\Psi}(\vec{k},\vec{k}')=
\int d\vec{r}d\vec{r}'\,e^{i\vec{k}\cdot\vec{r}}e^{i\vec{k}'\cdot\vec{r}'}\,
\Psi(\vec{r},\vec{r}').
\end{equation}
Notice that there is a factor $i^l$ in the multipole decomposition of 
$e^{i\vec{k}\cdot\vec{r}}$, and that 
$\left(i^l\right)^2$ is +1 for even values of $l$ and 
$-1$ for odd values of $l$. 
This leads to \cite{HS14}
\begin{equation}
\widetilde{\Psi}(\vec{k},\vec{k}')=\alpha\, 
\widetilde{\Psi}_{\rm ee}(\vec{k},\vec{k}')
-\beta\, \widetilde{\Psi}_{\rm oo}(\vec{k},\vec{k}'). 
\end{equation}
If one constructs a two-particle density in the momentum space with this 
wave function, the interference term therefore 
acts in the opposite way to that 
in the coordinate space. That is, the two-particle density in the 
momentum space is hindered for $\vec{k}\sim\vec{k}'$, while it is enhanced 
for $\vec{k}\sim-\vec{k}'$. 
This could be intuitively understood also from a point of view of uncertainty 
relation between coordinate and momentum. 

\begin{figure}[b]
\begin{center}
\includegraphics[clip,width=7.5cm]{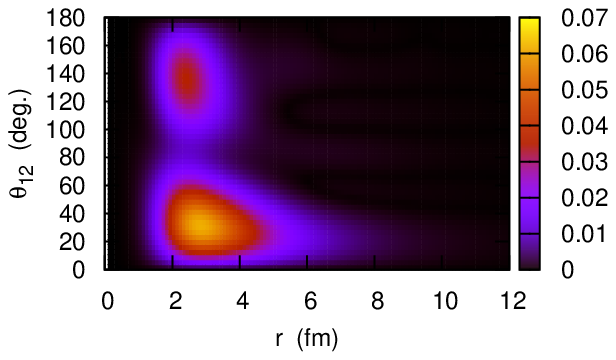}
\includegraphics[clip,width=7.5cm]{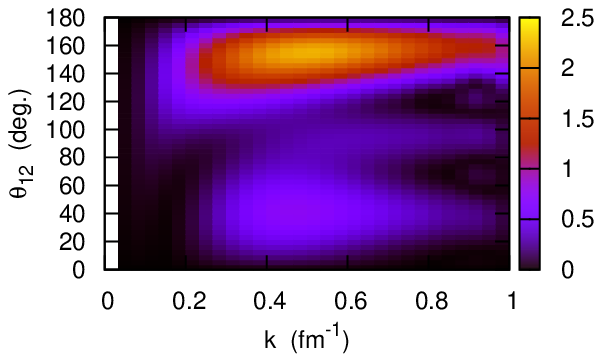}
\end{center}
\caption{The two-particle density of the $^6$He nucleus in the coordinate 
space (the left panel) and in the momentum space (the right panel). 
The density in the coordinate space is plotted as a function of 
$|\vec{r}|=|\vec{r}'|=r$ and the angle between 
$\vec{r}$ and $\vec{r}'$, while density 
in the momentum space is plotted as a function of $|\vec{k}|=|\vec{k}'|=k$ 
and the angle between $\vec{k}$ and $\vec{k}'$. Weight factors of 
8$\pi^2 r^4\sin\theta$ and 8$\pi^2 k^4\sin\theta$ 
are multiplied to the densities, and thus the units in the plots are fm$^{-2}$ 
and fm$^2$ for the left and the right panels, respectively. 
}
\end{figure}

Figure 1 shows the two-particle density for the $^6$He nucleus 
in the coordinate 
space (the left panel) and in the momentum space (the right panel), obtained 
with the three-body model of Ref. \cite{HS05}. 
To plot the density, we set $|\vec{r}|=|\vec{r}'|$ and 
$|\vec{k}|=|\vec{k}'|$. 
One can see that the density in the coordinate space is enhanced for small 
values of the angle between $\vec{r}$ and $\vec{r}'$. In contrast, 
the density in the momentum space is enhanced for large values of 
the angle between $\vec{k}$ and $\vec{k}'$. 

This leads to an interesting idea on two-nucleon emission. That is, two 
nucleons are confined inside a nucleus as 
a spatially compact dinucleon-like cluster, and they are 
thus first emitted together in a similar direction. 
Outside a nucleus, since the two nucleons 
do not bound, they move 
according to the momentum distribution which 
they originally have before the emission, 
that is,  the two nucleons fly apart in the opposite 
direction. This behavior has actually been seen in a recent 
time-dependent three-body model calculation for two-proton decay of 
$^6$Be \cite{OHS14}. 

\section{Two-neutron decay of $^{26}$O}

Let us now apply the three-body model to the two-neutron decay of 
$^{26}$O and discuss how the dineutron correlation 
affects its dynamics. 
To this end, we use 
a Woods-Saxon potential between a valence neutron and the core 
nucleus, $^{24}$O, in which the parameters are chosen in order to
reproduce the energy of the $d_{3/2}$ resonance state of $^{25}$O
at 770 keV \cite{H08}. 
For the two-body pairing interaction between the valence 
neutrons, we employ
a density-dependent contact interaction \cite{HS05}, whose parameters 
are adjusted to reproduce the ground state energy of $^{27}$F using 
a similar three body model of $^{25}$F+$n$+$n$. 

\subsection{Decay energy spectrum}

We first discuss the decay energy spectrum. 
With the three-body model for $^{26}$O, we compute it 
for a given angular momentum $I$ as,
\begin{equation}
\frac{dP_I}{dE}=\sum_k|\langle\Psi_k^{(I)}|\Phi^{(I)}_{\rm ref}\rangle|^2
\,\delta(E-E_k),
\label{decayspectrum}
\end{equation}
where $\Psi_k^{(I)}$ is a
solution of the three-body model Hamiltonian with
the angular momentum $I$ and the energy $E_k$,
and $\Phi^{(I)}_{\rm ref}$ is the wave function
for a reference state with the same angular momentum.
For the reference state, 
we use the uncorrelated two-neutron state
in $^{27}$F with the
$|[1d_{3/2}\otimes1d_{3/2}]^{(IM)}\rangle$ configuration, which is
dominant in the ground state of $^{27}$F, since $^{26}$O was produced 
in the experiment of Ref. \cite{Lunderberg12} 
with the proton knockout reaction of $^{27}$F. 
The actual calculations for the decay energy spectrum are done using 
the Green's function technique \cite{EB92}, by fully taking 
into account the continuum effects. See Refs. \cite{HS14,HS14b} for 
details. 

\begin{figure}[htb]
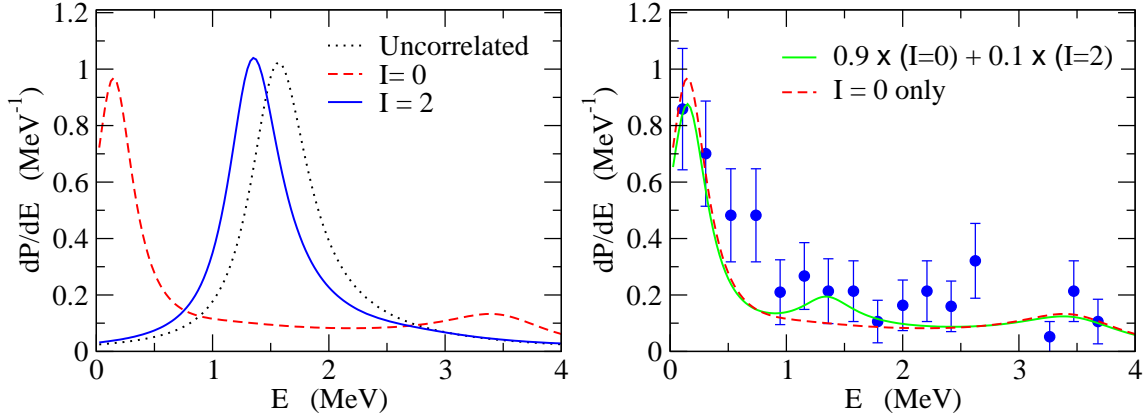

\begin{center}
\includegraphics[clip,width=7.5cm]{fig2a.eps}
\includegraphics[clip,width=7.5cm]{fig2b.eps}
\end{center}
\caption{
(the left panel)
The decay energy spectrum for the two-neutron emission decay of $^{26}$O.
The dashed and the solid lines are for the 0$^+$ and 2$^+$ states,
respectively. The dotted line shows the uncorrelated spectrum obtained
by ignoring the interaction between the valence neutrons.
(the right panel)
The decay energy spectrum obtained by superposing the $I=0$ and $I=2$
components as is indicated in the figure. The dashed line is the same
as the one in the left panel, that is, the decay energy spectrum for the
pure $I=0$ configuration.
The experimental data, normalized to the unit area, are
taken from Ref. \cite{Lunderberg12}.
}
\end{figure}

The left panel of Fig. 2 shows the decay energy spectrum
of $^{26}$O
for the $I=0$ (the dashed line) and $I$=2 (the solid line).
For a presentation purpose, we have introduced 
a width of 0.21 MeV to the spectrum. 
For comparison,
we also show the spectrum for the uncorrelated case by the dotted line,
which gives the same spectrum both for $I=0$ and $I=2$.
For the uncorrelated case, the spectrum has a peak at $E=1.54$ MeV, that
is twice the single-particle resonance energy, 0.77 MeV.
With the pairing interaction between the valence neutrons, the peak
energy is shifted towards lower energies. The energy shift is larger in $I=0$
than in $I=2$. That is, the peak in the spectrum appears at
$E=0.148$ MeV ($\Delta E= -1.392$ MeV) for
$I=0$ and at $E=1.354$ MeV ($\Delta E=-0.186$ MeV)
for $I=2$.

The fact that the 2$^+$ state appears at
an energy slightly smaller than the unperturbed energy 
is a natural consequence of 
the three-body model. 
In  standard textbooks of nuclear
physics, 
it is shown that the energy shift due to a pairing interaction,
$v(\vec{r}_1,\vec{r}_2)=-g \,\delta (\vec{r}_1-\vec{r}_2)$, is
evaluated for a single-$j$ orbit as,
\begin{equation}
\Delta E_I = \langle [jj]^{(IM)}|
-g \,\delta (\vec{r}_1-\vec{r}_2) |[jj]^{(IM)}\rangle 
=-gF_r\,\frac{(2j+1)^2}{8\pi}
\left(
\begin{array}{ccc}
j & j & I \\
1/2 & -1/2 & 0
\end{array}
\right)^2,
\end{equation}
where $F_r$ is the radial integral of the
four single-particle wave functions.
If one applies this formula to the $^{26}$O nucleus and sets $j=d_{3/2}$,
one obtains
\begin{equation}
\Delta E_{I=0}=-\frac{16}{8\pi}gF_r\cdot \frac{1}{4},~~~~~~
\Delta E_{I=2}=-\frac{16}{8\pi}gF_r\cdot \frac{1}{20}. 
\label{deltaE0}
\end{equation}
This equation predicts $\Delta E_{I=2}=-0.278$ MeV for $\Delta E_{I=0}=-1.392$
MeV, that is, $\Delta E_{I=0}/\Delta E_{I=2}$ =5. 
This value is compared to the calculated value
of $\Delta E_{I=0}/\Delta E_{I=2}$ =7.48 obtained with the present
three-body model. 
Even though the ratio 
$\Delta E_{I=0}/\Delta E_{I=2}$ in the three-body model somewhat deviates 
from the simple estimates
of Eq. (\ref{deltaE0}) 
due to the many-body continuum  effects,
the small energy shift for the 2$^+$ state
can be well understood by these formulas derived for
the single-$j$ model with the residual pairing interaction.
Notice that the 2$^+$ energy never exceeds the unperturbed
energy, and thus it must be smaller than 1.54 MeV
for the $^{26}$O nucleus. 

The right panel of Fig. 2 shows the decay energy spectrum obtained by
including the contribution of the $I=2$ configuration. 
To this end, we take a linear combination of 
the $I=0$ and $I=2$ contributions,
that is,
\begin{equation}
\frac{dP}{dE}=(1-\gamma)\frac{dP_{I=0}}{dE}+\gamma\frac{dP_{I=2}}{dE}.
\end{equation}
The actual value of $\gamma$ would
depend on the details of the wave function of $^{27}$F as well as
the reaction dynamics of the proton knock-out reaction of $^{27}$F
with which the
initial state of $^{26}$O was prepared in the experiment
of Ref. \cite{Lunderberg12}.
Here we
arbitrarily take $\gamma$=0.1.
For comparison,
the figure also shows
the pure $I=0$ component, which is the same as that shown in the left panel.
One can see that the experimental data are
reproduced slightly better by
mixing the $I=2$ component, although the error bars are large and one may not
draw a definite conclusion.

We mention that we obtain 
the 2$^+$ state of $E=1.338$ MeV if we use the pairing interaction which 
yields the ground state energy of 5 keV rather than 148 keV. 
This value is similar to the previous result,
$E=1.354$ MeV, and we thus conclude that the energy of the 2$^+$
state is much less sensitive to the $nn$ interaction as compared
to the 0$^+$ state.
This implies that the 2$^+$ state of $^{26}$O should definitely
appear around $E=1.3$ MeV as long as the three-body picture is correct.

\subsection{Angular distribution of the emitted neutrons}

Let us next discuss the angular distribution of the emitted neutrons. 
The dotted line in Fig. 3 shows 
the distribution obtained without including the 
$nn$ interaction, which 
is almost symmetric
around $\theta_{12}=\pi/2$.
In the presence of the $nn$ interaction, the angular distribution
becomes highly asymmetric, in which the emission of two neutrons in the
opposite direction (that is, $\theta_{12}=\pi$) is
enhanced, as is shown
by the solid line.
The average angle is calculated to be
$\langle \theta_{12}\rangle = 115.3^\circ$.
This behavior should reflect properties of the resonance wave 
function of $^{26}$O, as we have discussed in Sec. 2. 
That is, because of the continuum couplings, several configurations with 
opposite parity states mix coherently, which leads to an enhancement 
of $\theta\sim 0$ in the coordinate space and of 
$\theta\sim \pi$ in the momentum space (for the $^{26}$O nucleus, 
it is mainly due to the interference between the
$l=0$ and $l=1$
components, as higher partial waves are considerably 
suppressed inside the 
centrifugal barrier 
for $e_1\sim e_2 \sim E_{\rm g.s.}/2$=0.07 MeV). See also Ref. \cite{Grigorenko}. 
We can therefore conclude that, if an enhancement in the region 
of $\theta\sim\pi$ in the angular distribution was observed experimentally, that would 
make a clear evidence for the dineutron correlation in this nucleus. 

\begin{figure}[bt]
\begin{center}
\includegraphics[width=7.5cm]{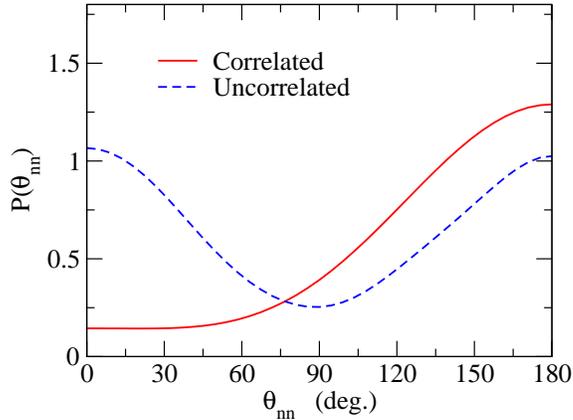}
\end{center}
\caption{
The probability distribution with respect to the 
opening angle of the momentum vectors of the emitted
two neutrons from
 $^{26}$O.
The solid and the dashed lines show the correlated and uncorrelated results,
respectively. }
\end{figure}

\section{Summary}

We have discussed a role of dineutron correlation in 
the two-neutron emission decay of the
neutron unbound nucleus $^{26}$O. 
To this end, we have 
used the three-body model with a contact neutron-neutron ($nn$) 
interaction. 
In the absence of the $nn$ interaction, the two valence neutrons 
occupy the $d_{3/2}$ resonance state at 0.77 MeV in $^{25}$O and thus 
the unperturbed energy for $^{26}$O is 1.54 MeV above the $2n$ threshold. 
In the presence of the $nn$ interaction, the energy is shifted towards 
low energies. For the ground state, 
the experimental data suggest that the energy is shifted almost to the 
threshold. On the other hand, 
we have shown that the 2$^+$ state
should appear at around $E=1.35$ MeV and this value does not change much
even if we change the $nn$ interaction to vary the ground state energy
from 150 keV to 5 keV.
This 2$^+$ energy is close to, but slightly smaller than,
the unperturbed energy and
thus the energy shift from the unperturbed energy is much smaller
than the energy shift for the 0$^+$ state.
We have argued that this
is a typical spectrum
well understood by the single-$j$ model with the 
pairing residual interaction.

We have also discussed that 
the density distribution 
of a three-body resonance state in $^{26}$O 
is strongly reflected in
the angular distribution of the emitted neutrons.
In particular,
the emission of the two neutrons in back-to-back angles is 
enhanced in the angular distribution,
that can be interpreted as a clear evidence for the dineutron correlation.
That is, the coherent admixture of many configurations with opposite parity states 
leads to the dineutron correlation, with which the 
two valence neutrons tend 
to have momenta in the opposite directions. 

So far, the angular distribution for the two-neutron decay
of $^{26}$O has not yet been measured experimentally.
It would be extremely intriguing if it will be measured at
new generation RI beam facilities, such as
the SAMURAI facility at RIBF at RIKEN \cite{AN13}.
A measurement of the 2$^+$ state would also be useful in order to 
understand the ground state properties of the unbound $^{26}$O nucleus. 

\section*{Acknowledgments}

We thank T. Oishi, Y. Kondo, T. Nakamura, and Z. Kohley for useful discussions.
This work was supported by
JSPS KAKENHI Grant Numbers
25105503 and 26400263.

\end{document}